\documentclass{PoS}

\def\d{{\rm d}}
\def\mom#1{\langle #1 \rangle}

\title{Determining $\alpha_{s}$ at NNLO from Event Shape Data}

\ShortTitle{$\alpha_{s}$ from Event Shape Data}

\author{\speaker{Gionata Luisoni}\\
       Institut f\"ur Theoretische Physik, Universit\"at Z\"urich, CH-8057 Z\"urich, Switzerland\\
        E-mail: \email{luisonig@physik.uzh.ch}}

\abstract{We report on several methodologically different NNLO determinations of the strong coupling constant from event shapes and related observables in e+e- annihilation.
The event shape distributions are analyzed within a combined framework of NNLO and NNLA resummation. We also investigate the role of hadronisation corrections, using both Monte Carlo generator predictions and analytic models to parametrise non-perturbative power corrections.}

\FullConference{XVIII International Workshop on Deep-Inelastic Scattering and Related Subjects, DIS 2010\\
		April 19-23, 2010\\
		Firenze, Italy}

\begin{document}

\section{Introduction}\label{sec:intro}

The reaction of $e^{+}e^{-}$ annihilation into 3 jets has played historically a very prominent role for phenomenology and allows a precise determination of the strong coupling constant $\alpha_{s}$, since the deviation from two-jet configurations is proportional to it. The phenomenologically interesting observables for the purpose of determining the strong coupling constant can be divided into two big categories. On one side there are jet rates, which rely on a jet algorithm and are based on the number of jets observed in an event, on the other side also the topology of the single events can be studied in a systematic fashion by means of so-called event-shape observables. The popularity of the latter is mainly due to the fact that they are well suited both for experimental measurement and for theoretical description, since many of them are infrared and collinear safe. They describe topological properties of hadronic final states by parameterizing the energy-momentum flow of an event. This class of observables is also interesting for the study of hadronisation effects. Hadronisation corrections usually result in a distortion of the event-shape distributions and are thus not easily disentangled from partonic predictions. However, the study of mean values and higher moments of event shapes allows a clear separation of perturbative and non-perturbative predictions. Furthermore, the comparison of hadronization corrections computed by general purpose Monte Carlo programs with predictions from analytical models permits to shed some light on the effects of hadronization corrections in the determination of $\alpha_{s}$.

\begin{figure}
\centering
  \includegraphics[scale=0.44]{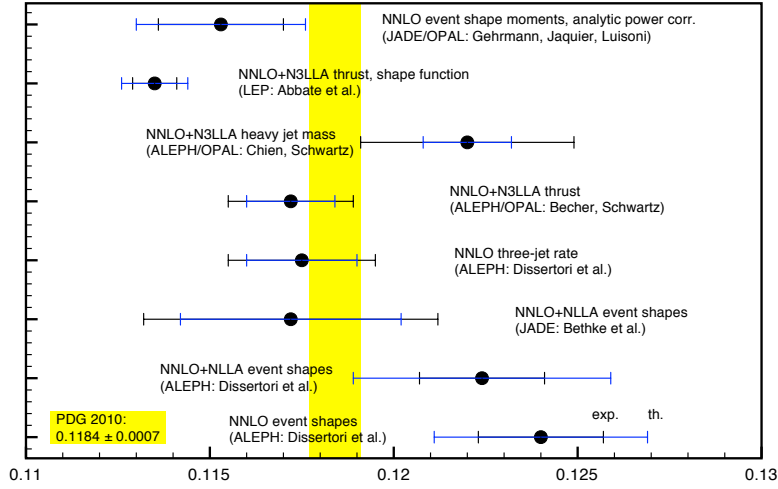}\\
  \vspace{-1.0cm}
  \caption{Recent determinations of $\alpha_s$ based NNLO predictions taken from Refs.~\cite{Dissertori:2007xa,Dissertori:2009ik,Gehrmann:2009eh,Dissertori:2009qa,Becher:2008cf}.}\label{fig:as}
\end{figure}

We have studied jet rates, the distributions and the first five moments of six event-shape observables: thrust $T$ (respectively, $\tau = 1-T$), heavy jet mass $\rho$, wide and total jet broadening $B_W$ and $B_T$, $C$-parameter and the two-to-three-jet transition parameter in the Durham algorithm, $Y_3$. For the definitions of these variables and their historical origin we refer to~\cite{Jones:2003yv} and references therein. We will denote the variables collectively as $y$ in the following, such that their two jet limit is $y\to 0$. Recently, with the appearance of the NNLO results, several extractions of $\alpha_{s}$ have been performed using event-shape observables~\cite{Dissertori:2007xa,Dissertori:2009ik,Gehrmann:2009eh,Dissertori:2009qa,Becher:2008cf}. Their results are summarized in Fig.~\ref{fig:as}. 
In the following we present three of them~\cite{Dissertori:2009ik,Dissertori:2009qa,Gehrmann:2009eh}.

\section{$\alpha_{s}$ from Event-Shape Distributions}

The fixed-order QCD description of event-shape distributions is given by a perturbative expansion of the form
\begin{eqnarray}
\frac{1}{\sigma_{\textrm{had}}}\, \frac{\d\sigma}{\d y}
(y,Q,\mu) &=& \bar\alpha_s (\mu) \frac{\d {A}}{\d y}(y)
+ \bar\alpha_s^2 (\mu) \frac{\d {B}}{\d y} (y,x_\mu) +
\bar\alpha_s^3 (\mu) \frac{\d {C}}{\d y}(y,x_\mu) +
{\cal O}(\bar\alpha_s^4)\;, \label{eq:NNLO0mu}
\end{eqnarray}
where $\bar\alpha_s = \frac{\alpha_s}{2\pi}$ and $x_\mu = \frac{\mu}{Q}$, and where $A$, $B$ and $C$ are the perturbatively calculated coefficients~\cite{GehrmannDeRidder:2007hr} at LO, NLO and NNLO. The distribution is normalised to the total hadronic cross section $\sigma_{\textrm{had}}$ in $e^+e^-$ annihilation. The dependence of~(\ref{eq:NNLO0mu}) on the collision energy is only through $\alpha_s$ and $x_\mu$ and the scale dependence of $\alpha_s$ is determined according to the three-loop running of $\alpha_{s}(\mu)$.

In order to obtain reliable predictions over the full kinematical range, the perturbative fixed-order prediction~(\ref{eq:NNLO0mu}) has to be matched with resummation, which is taken into account at next-to-leading logarithmic (NLL) accuracy in the $\ln\,R$-matching scheme. For more details about the NLLA+NNLO matching we refer to Ref.~\cite{Gehrmann:2008kh} and references therein.

We have used the six event-shape observables listed in Section~\ref{sec:intro} for our fits. The measurements we use have been carried out by the {\small ALEPH} collaboration \cite{aleph-qcdpaper} at eight different centre-of-mass (CM) energies between 91.2 and 206\,GeV. 
The perturbative QCD prediction is corrected for hadronisation and resonance decays by means of a transition matrix, which is computed with the MC generators {\small PYTHIA}~\cite{Sjostrand:2000wi}, {\small HERWIG}~\cite{Corcella:2000bw}  and {\small ARIADNE}~\cite{Lonnblad:1992tz}, all tuned to global hadronic observables at $M_Z$~\cite{Barate:1996fi}.
Corrected measurements of event-shape distributions are compared to the theoretical calculation at particle level. For a detailed description of the determination and treatment of experimental systematic uncertainties we refer to Refs.~\cite{aleph-qcdpaper,Dissertori:2007xa,Dissertori:2009ik}.

The combined results of six event-shape variables and eight LEP1/LEP2 CM energies is
\begin{center}
    $\alpha_s(M_Z) = 0.1224
    \;\pm\; 0.0009\,\mathrm{(stat)}
    \;\pm\; 0.0009\,\mathrm{(exp)}
    \;\pm\; 0.0012\,\mathrm{(had)}
    \;\pm\; 0.0035\,\mathrm{(theo)}\;.$
\end{center}
For the fitted values of the coupling constant as found from event-shape variables calculated at various orders we refer to the figures and tables of~\cite{Dissertori:2009ik}. The central value of the result is slightly lower than the central value of 0.1228 obtained from a fit using purely fixed-order NNLO predictions~\cite{Dissertori:2007xa}. Furthermore the dominant theoretical uncertainty on  $\alpha_s(M_Z)$, as estimated from scale variations, is reduced by 20\% compared to NLO+NLLA and the scatter among the values of $\alpha_s(M_Z)$ extracted from the six different event-shape variables is substantially reduced. However, compared to the fit based on purely fixed-order NNLO predictions, the perturbative uncertainty is {\it increased} in the NNLO+NLLA fit. The reason is that in the two-jet region the NLLA+NLO and NLLA+NNLO predictions agree by construction and therefore the renormalisation scale uncertainty is dominated by the resummation in this region, which results in a larger overall scale uncertainty in the $\alpha_s$ fit.

Apart from the $\alpha_{s}$ determination using the standard MC generators mentioned above, we used {\small HERWIG++}\,\cite{LatundeDada:2007jg} version 2.3 together with the {\small MCNLO} \cite{Frixione:2002ik} and {\small POWHEG} \cite{Nason:2004rx} schemes for investigating hadronization corrections.
From this study it appears that there are two ``classes'' of variables. With standard hadronisation corrections from {\small PYTHIA} we obtain $\alpha_s(M_Z)$ values some $5\%$ higher for the first class, consisting of $T$, $C$ and $B_T$, which still suffer from sizable missing higher order corrections, than for the second class consisting of the $\rho$, $B_W$ and $Y_3$, which have a better perturbative stability~\cite{GehrmannDeRidder:2009dp}.
This means that the {\small PYTHIA} hadronisation corrections, applied in the $\alpha_s$ fit, might be too small for the first class of variables, resulting in a larger $\alpha_s(M_Z)$ value. For further details of our analysis we refer to Ref.~\cite{Dissertori:2009ik}.

\section{$\alpha_{s}$ from Moments of Event Shapes}

The $n$th moment of an event-shape observable $y$ is defined by
\begin{equation}
\mom{y^n}=\frac{1}{\sigma_{\rm{had}}}\,\int_0^{y_{\rm{max}}} y^n
 \frac{\d\sigma}{\d y} \d y \;,
\label{eq:momdef}
\end{equation}
where $y_{\mathrm{max}}$ is the kinematically allowed upper limit of the observable.
For moments of event shapes, one expects the hadronisation corrections to be additive, such that the cross section can be divided into a perturbative and a non-perturbative contribution, where the non-perturbative contribution accounts for hadronisation effects.

In ref.~\cite{Gehrmann:2009eh}, the dispersive model derived in Refs.~\cite{Dokshitzer:1995qm,Dokshitzer:1998pt,Dokshitzer:1998qp} has been used and extended to NNLO to estimate hadronisation corrections to event-shape moments by calculating analytical predictions for power corrections. It introduces only a single new parameter $\alpha_0$, which can be interpreted as the average strong coupling in the non-perturbative region:
\begin{equation}
\frac{1}{\mu_I}\int_0^{\mu_I} dQ \,\alpha_{\rm{eff}}(Q^2)=\alpha_0(\mu_I)\;,
\label{alpha0}
\end{equation}
where below the IR cutoff $\mu_{I}$ the strong coupling is replaced by an effective coupling. This dispersive model for the strong coupling leads to a shift in the distributions
\begin{equation}
\frac{\d\sigma}{\d y}(y)=\frac{\d\sigma_{\rm{pt}}}{\d y}\,(y-a_y\,P)\;,
\label{eq:disp}
\end{equation}
where the numerical factor $a_y$ depends on the event shape, while ${P}$ is believed to be universal and scales with the CM energy like $\mu_I/Q$. Insertion of eq. (\ref{eq:disp}) into~(\ref{eq:momdef}) leads to
\begin{equation}
\qquad\langle y^n \rangle = \int^{y_\mathrm{max}-a_yP}_{-a_yP}\mathrm{d}y\,(y+a_yP)^n\frac{1}{\sigma _\mathrm{tot}}\frac{\mathrm{d}\sigma_\mathrm{pt}}{\mathrm{d}y}(y)
\approx \int^{y_\mathrm{max}}_0\mathrm{d}y\,(y+a_yP)^n\frac{1}{\sigma _\mathrm{tot}}\frac{\mathrm{d}\sigma_\mathrm{pt}}{\mathrm{d}y}(y)\;.
\end{equation}
From this expression one can extract the non-perturbative predictions for the moments of $y$.

 The expressions derived in~\cite{Gehrmann:2009eh} match the dispersive model with the perturbative prediction at NNLO QCD. Comparing these expressions with experimental data on event-shape moments, a combined determination of the perturbative strong coupling constant $\alpha_s$ and the non-perturbative parameter $\alpha_0$ has been performed~\cite{Gehrmann:2009eh}, based on data from the {\small JADE} and {\small OPAL} experiments~\cite{Pahl:2008uc}. The data consist of 18 points at CM energies between 14.0 and 206.6 GeV for the first five moments of $T$, $C$, $Y_3$, $\rho$, $B_W$ and $B_T$, and have been taken from \cite{Pahl:2007zz}. For each moment the NLO as well as the NNLO prediction was fitted with $\alpha_s(M_Z)$ and $\alpha_0$ as fit parameters, except for the moments of $Y_3$, which have no leading $\frac{1}{Q}$ power correction and thus are independent of $\alpha_0$.

Compared to previous results at NLO, inclusion of NNLO effects results in a considerably improved consistency in the parameters determined from different shape variables, and in a substantial reduction of the error on $\alpha_s$.
Furthermore the theoretical error on the extraction of $\alpha_S(M_Z)$ from $\rho$, $Y_3$ and $B_W$ is considerably smaller than from $\tau$, $C$ and $B_T$. As mentioned above, the moments of the former three shape variables receive moderate NNLO corrections for all $n$, while the NNLO corrections for the latter three are large already for $n=1$ and increase with $n$. Consequently, the theoretical description of the moments of $\rho$, $Y_3$ and $B_W$ displays a higher perturbative stability, which is reflected in the smaller theoretical uncertainty on  $\alpha_s(M_Z)$ derived from those variables.

In a second step, we combine the $\alpha_s(M_Z)$ and $\alpha_0$ measurements obtained from different event-shape variables. Taking the weighted mean over all values except $B_W$ and $B_T$,  we obtain at NNLO:
$$\alpha_s(M_Z)=0.1153\pm0.0017(\mathrm{exp})\pm0.0023(\mathrm{th})\,, \quad \alpha_0=0.5132\pm0.0115(\mathrm{exp})\pm0.0381(\mathrm{th})\,.$$
The moments of $B_W$ and $B_T$ have been excluded here since their theoretical description requires an additional contribution to the non-perturbative coefficient $P$~\cite{Gehrmann:2009eh} which is unknown to NNLO.

The average of $\alpha_s(M_Z)$ is dominated by the measurements based on $\rho$ and $Y_3$, which have the smallest theoretical uncertainties. From NLO to NNLO~\cite{Gehrmann:2009eh}, the error on $\alpha_s(M_Z)$ is reduced by a factor of two.
The error on $\alpha_s(M_Z)$ is clearly dominated by the $x_{\mu}$ variation, while the largest contribution to the error on $\alpha_0$ comes from the uncertainty on the Milan factor ${\cal M}$~\cite{Dokshitzer:1998pt}, which has not been improved in the current study.

To quantify the difference of the dispersive model to hadronisation corrections from the legacy generators, we analysed the moments of $(1-T)$ with hadronisation corrections from {\small PYTHIA}. As a result, we obtained fit results for $\alpha_s(M_{\mathrm Z})$ which are typically 4\% higher than by using the dispersive model, with a slightly worse quality of the fit. Comparing perturbative and non-perturbative contributions at $\sqrt{s} = M_{{\mathrm Z}}$, we observed that {\small PYTHIA} hadronisation corrections amount to less than half the power corrections obtained in the dispersive model, thereby explaining the tendency towards a larger value of $\alpha_s(M_{\mathrm Z})$, since the missing numerical magnitude of the power corrections must be compensated by a larger perturbative contribution.

\section{$\alpha_{s}$ from Jet Rates}

Very recently a new determination of $\alpha_{s}$ was performed using jet rates~\cite{Dissertori:2009qa}. Theoretical NNLO predictions for jet rates~\cite{GehrmannDeRidder:2008ug} as a function of the jet resolution parameter $y_{\rm cut}$ are compared to ALEPH data~\cite{aleph-qcdpaper} using the Durham jet algorithm, for which the distance measure is given by
\begin{equation}
y_{ij,D}\,=\,2\,\min(E_{i}^{2},\,E_{j}^{2})\,\left(1\,-\,\cos\theta_{ij}\right)/E_{\textrm{vis}}^{2}\,
\end{equation}
where $E_{\textrm{vis}}$ denotes the energy sum of all particles in the final state. In Ref.~\cite{GehrmannDeRidder:2008ug} it was shown that NNLO predictions for jet rates have only very small hadronization corrections and the theoretical error for $10^{-1}<y_{\rm cut}<10^{-2}$ drops below the per-cent relative uncertainty. This motivates a dedicated extraction of $\alpha_{s}$. The corrected ALEPH measurements for the three-jet rate are compared to the theoretical calculation at particle level. Values for $\alpha_{s}(M_Z)$ are obtained by a least-squares fit, performed separately for each $y_{\rm cut}$ value in the range . A nice stability of the result is found up to values of $\ln y_{\rm cut}\approx -4.5$. As final result the value for $y_{\rm cut}=0.02$ is taken, which represents an optimal compromise between minimal systematic uncertainty and stability. The following values for $\alpha_{s}$ is found:
\begin{equation}
\alpha_s(M_Z)=0.1175\pm0.0020(\mathrm{exp})\pm0.0015(\mathrm{th}),\nonumber\,.
\end{equation}
Results from LEP2 energies give similar central values but larger statistical uncertainties. Combining the errors in quadrature yields $\alpha_s(M_Z)=0.1175\pm0.0025$, which is in excellent agreement with the latest world average~\cite{Bethke:2009jm}.
This verifies the expectations that the three-jet rate is an excellent observable for this kind of analysis, thanks to the good behaviour of its perturbative and non-perturbative contributions over a sizable range of jet-resolution parameters.

\section{Acknowledgments}
This research was supported in part by the Swiss National Science Foundation (SNF) under contracts PP0022-118864 and 200020-126691.

\end{document}